\documentclass[
prl,aps,
reprint,
a4paper,
superscriptaddress,
longbibliography,
floatfix
]{revtex4-1}
\usepackage{graphicx}
\usepackage{epsfig}
\usepackage{amsfonts}
\usepackage{amsmath}
\usepackage{amsthm}
\usepackage{amssymb}
\usepackage{color}
\usepackage{multirow}
\usepackage[colorlinks=true,linkcolor=blue,citecolor=magenta,urlcolor=blue]{hyperref}

\usepackage{graphicx}
\usepackage[caption=false,font=large]{subfig}

\newtheorem{theorem}{Theorem}

\newcommand{\ket}[1]{|#1\rangle}

\begin{document}


\title{Optimal Classical Simulation of State-Independent Quantum Contextuality}



\author{Ad\'an Cabello}
\email{adan@us.es}
\affiliation{Departamento de F\'{\i}sica Aplicada II, Universidad de Sevilla, E-41012 Sevilla, Spain}

\author{Mile Gu}
\affiliation{School of Physical and Mathematical Sciences, Nanyang Technological University, 21 Nanyang Link, Singapore 637371, Singapore}
\affiliation{Complexity Institute, Nanyang Technological University, 18 Nanyang Drive, Singapore 637723, Singapore}
\affiliation{Centre for Quantum Technologies, National University of Singapore, 3 Science Drive 2, Singapore 117543, Singapore}

\author{Otfried G\"uhne}
\affiliation{Naturwissenschaftlich-Technische Fakult\"at, Universit\"at Siegen,
Walter-Flex-Stra{\ss}e 3, D-57068 Siegen, Germany}

\author{Zhen-Peng Xu}
\affiliation{Theoretical Physics Division, Chern Institute of Mathematics,
	Nankai University,
	Tianjin 300071, People's Republic of China}
\affiliation{Departamento de F\'{\i}sica Aplicada II,
	Universidad de Sevilla,
	E-41012 Sevilla, Spain}


\begin{abstract}
Simulating quantum contextuality with classical systems requires memory. A fundamental yet open question is what is the minimum memory needed and, therefore, the precise sense in which quantum systems outperform classical ones. Here, we make rigorous the notion of classically simulating quantum state-independent contextuality (QSIC) in the case of a single quantum system submitted to an infinite sequence of measurements randomly chosen from a finite QSIC set. We obtain the minimum memory needed to simulate arbitrary QSIC sets via classical systems under the assumption that the simulation should not contain any oracular information. In particular, we show that, while classically simulating two qubits tested with the Peres-Mermin set requires $\log_2 24 \approx 4.585$ bits, simulating a single qutrit tested with the Yu-Oh set requires, at least, $5.740$ bits.
\end{abstract}



\maketitle


{\em Introduction.---}Quantifying the resources needed to simulate quantum 
phenomena with classical systems is crucial to making precise the sense in 
which quantum systems provide an advantage over classical ones. While the 
extra resources needed for simulating entanglement and quantum nonlocality 
(i.e., the quantum violation of Bell inequalities \cite{Bell64}) have been 
studied extensively \cite{BCT99,Steiner00,CGM00,BT03,TB03,Pironio03,CGMP05}, 
the resources needed to simulate quantum contextuality \cite{Bell66,KS67}, 
a natural generalization of quantum nonlocality to the case of nonspacelike 
separated systems and witnessed by the quantum violation of noncontextuality 
inequalities \cite{KCBS08,Cabello08,BBCP09,YO12,KBLGC12}, have been less 
explored \cite{K11,Cabello12,FK16}. In a nutshell, while simulating quantum 
nonlocality with classical systems requires superluminal communication \cite{BCT99,BT03,TB03,Pironio03,CGMP05}, simulating quantum contextuality 
requires memory \cite{K11,Cabello12,FK16} or, more precisely, the ability 
of storing and recovering a certain amount of classical information. It is 
known that, in some cases, the required memory is larger than the 
information-carrying capacity of the corresponding quantum system \cite{K11}. 
The problem is that only lower bounds to the minimum memory are known for 
some particular scenarios \cite{K11,FK16}. In addition, it is not known 
how the minimum memory scales with, e.g., the size of the set of possible 
measurements.


\begin{figure}[t]
\includegraphics[trim = 5.2cm 0.8cm 5.0cm 21.0cm,clip,width=8.0cm]{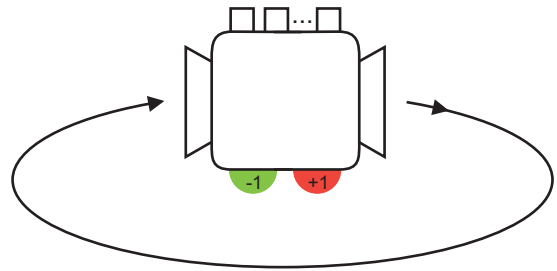}
	\caption{\label{Fig1} Contextuality experiment on a recycled system. Buttons represent possible measurements. Light bulbs represent possible outcomes. We consider experiments in which there are as many buttons as elements of the QSIC set, and all of them have two outcomes.}
\end{figure}


A particularly interesting case is that of quantum state-independent contextuality 
(QSIC) in experiments with sequential measurements \cite{Cabello08,BBCP09,YO12,KBLGC12} 
on a single recycled quantum system \cite{K11,WLK16,LMZNACH17}. In this case, a 
single quantum system is submitted to an unlimited sequence of measurements, 
randomly chosen from a finite set of measurements, as illustrated in 
Fig.\ \ref{Fig1}. After each measurement, the outcome is observed and recorded.
The set of measurements has the peculiarity of being able to 
produce contextuality no matter what the initial quantum state of the system is. 
These sets are called QSIC sets \cite{CKB15,CKP16} and, for each of them, there 
are optimal combinations of correlations for detecting contextuality \cite{KBLGC12}. 
The interest of this case comes from the fact that unbounded strings of data with 
contextual correlations can be produced using a {\em single} system initially 
prepared in an arbitrary state \cite{LMZNACH17}, a situation that strongly 
contrasts with the case of nonlocality generated through the violation of a Bell
inequality, where thousands of spacelike separated pairs of quantum systems 
in an entangled quantum state are needed. The question we want to address in 
this Letter is what is the minimal amount of memory a classical system would 
need to simulate the predictions of quantum theory for QSIC experiments 
with unlimited sequential measurements. Contrary to the previous approaches
\cite{K11, FK16}, we aim at simulating all statistics arising in quantum theory
and not only the perfect correlations leading to a violation of a contextuality
inequality. We consider the most general simulation under the restriction that 
the classical model used for simulation should not contain oracular information, 
as explained below.


{\em Scenario.---}We consider ideal experiments in which successive measurements 
are performed on a single quantum system at times $t_1<t_2<\cdots$ At each $t_i$, 
a measurement belonging to a QSIC set is randomly chosen and performed. We assume 
that the quantum state after the measurement at $t_i$ is the quantum state before 
the measurement at $t_{i+1}$. The process is repeated infinitely many times. Our 
aim is to extract conclusions valid for any QSIC set. However, the sake of clarity, 
we will present our results using two famous QSIC sets.


\begin{figure}[t]
	\centering
	\includegraphics[width=5.6cm]{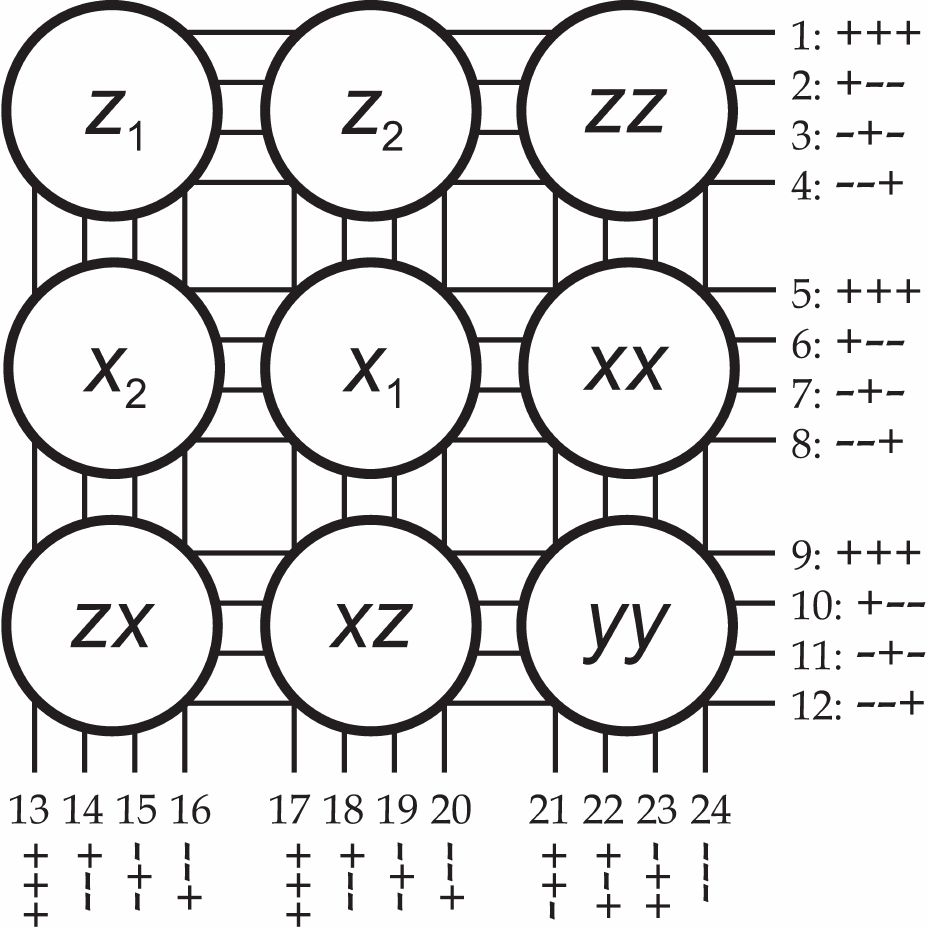}
	\caption{Observables in the Peres-Mermin set. $z_1$ denotes the quantum observable represented by the operator $\sigma_z^{(1)} \otimes \openone^{(2)}$. Similarly, $zx$ denotes $\sigma_z^{(1)} \otimes \sigma_x^{(2)}$. Observables in each row or column are mutually compatible and their corresponding operators have four common eigenstates. In the figure these eigenstates are represented by straight lines numbered from $1$ to $24$. For example, quantum state $|2\rangle$ is the one satisfying $\sigma_z^{(1)} \otimes \openone^{(2)} |2\rangle=|2\rangle$, $\openone^{(1)} \otimes \sigma_z^{(2)} |2\rangle=-|2\rangle$, and $\sigma_z^{(1)} \otimes \sigma_z^{(2)} |2\rangle=-|2\rangle$. \label{Fig2}}
\end{figure}


{\em The Peres-Mermin set.---}The QSIC set with the smallest number of observables 
known has nine 2-qubit observables and it is shown in Fig.~\ref{Fig2}. It was 
introduced by Peres \cite{Peres90} and Mermin \cite{Mermin90b} and first implemented 
in experiments with sequential measurements by Kirchmair {\em et al.} \cite{KZG09} 
on trapped ions and by Amselem {\em et al.} \cite{ARBC09} on single photons. In addition, 
it has been recently implemented on entangled photons by Liu {\em et al.} \cite{LHC16}.

When one uses this set for unlimited sequential measurements on a single system, 
from the moment two different observables that are in the same row or column in 
Fig.~\ref{Fig2} are measured consecutively, the system remains in one of the 24 
quantum states defined in Fig.~\ref{Fig2}. After that, any other subsequent 
measurement leaves the system in one of these 24 quantum states and each of 
them occurs with the same probability.


{\em The Yu-Oh set.---}As proven in Ref.~\cite{CKP16}, the QSIC set with the smallest
number of observables represented by rank-one projectors has 13 single-qutrit observables. 
It was introduced by Yu and Oh \cite{YO12} and is a subset of a QSIC set previously 
considered by Peres \cite{Peres91}. Its associated optimal noncontextuality inequality 
was found by Kleinmann {\em et al.} \cite{KBLGC12}. It inspired a photonic experiment
by Zu {\em et al.} \cite{ZWDCLHYD12} (see also Amselem {\em et al.}~\cite{ABB13}) and was implemented 
as an experiment with sequential measurements on a single ion by Zhang {\em et al.} 
\cite{ZUZ13}, and, recently, it was used to implement the scheme in Fig.\ \ref{Fig1} by 
Leupold {\em et al.} \cite{LMZNACH17}.

\begin{figure}
	\begin{tabular}{cccc}
		\subfloat[]{\includegraphics[width=0.2\textwidth]{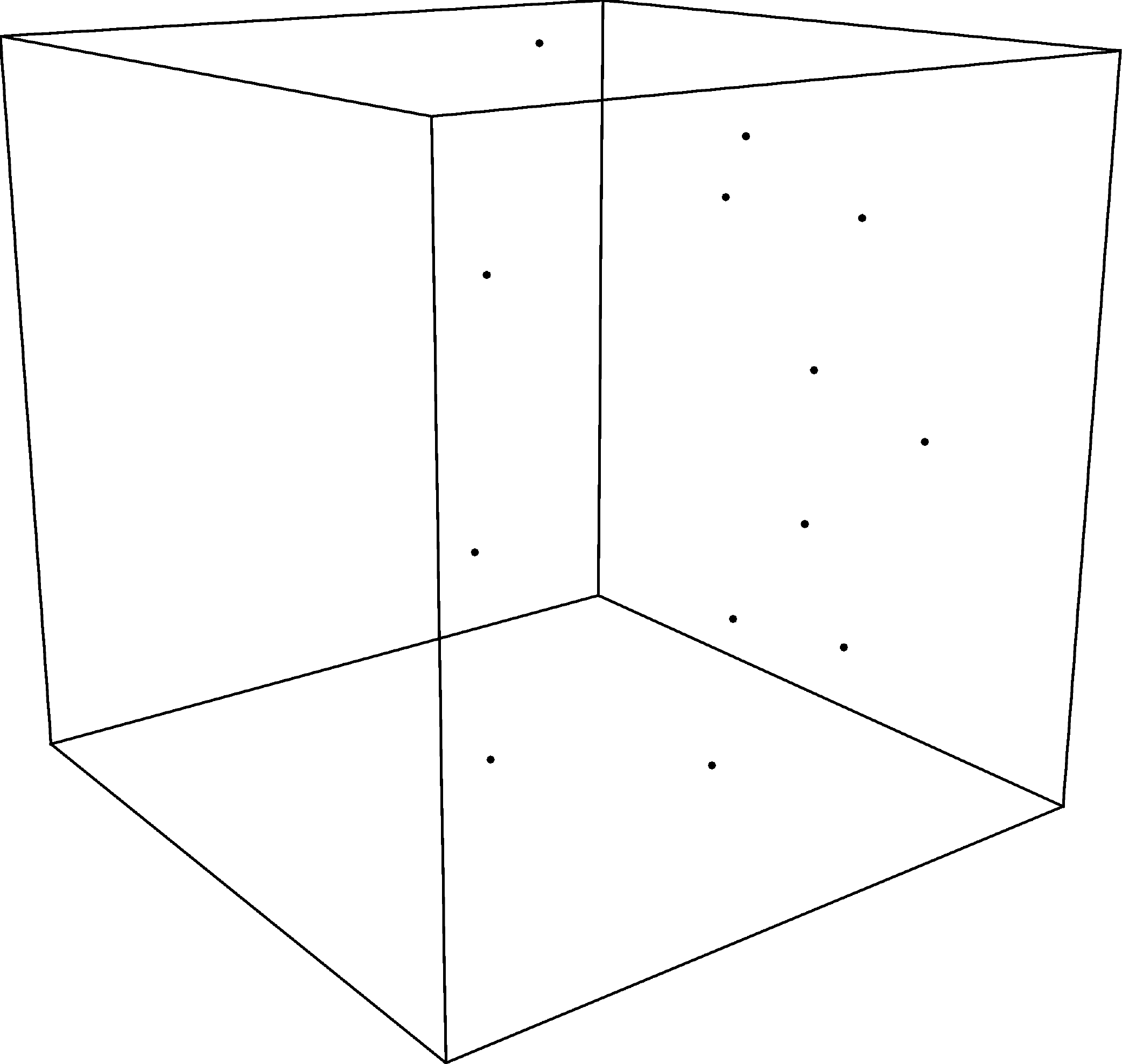}} & \subfloat[]{\includegraphics[width=0.2\textwidth]{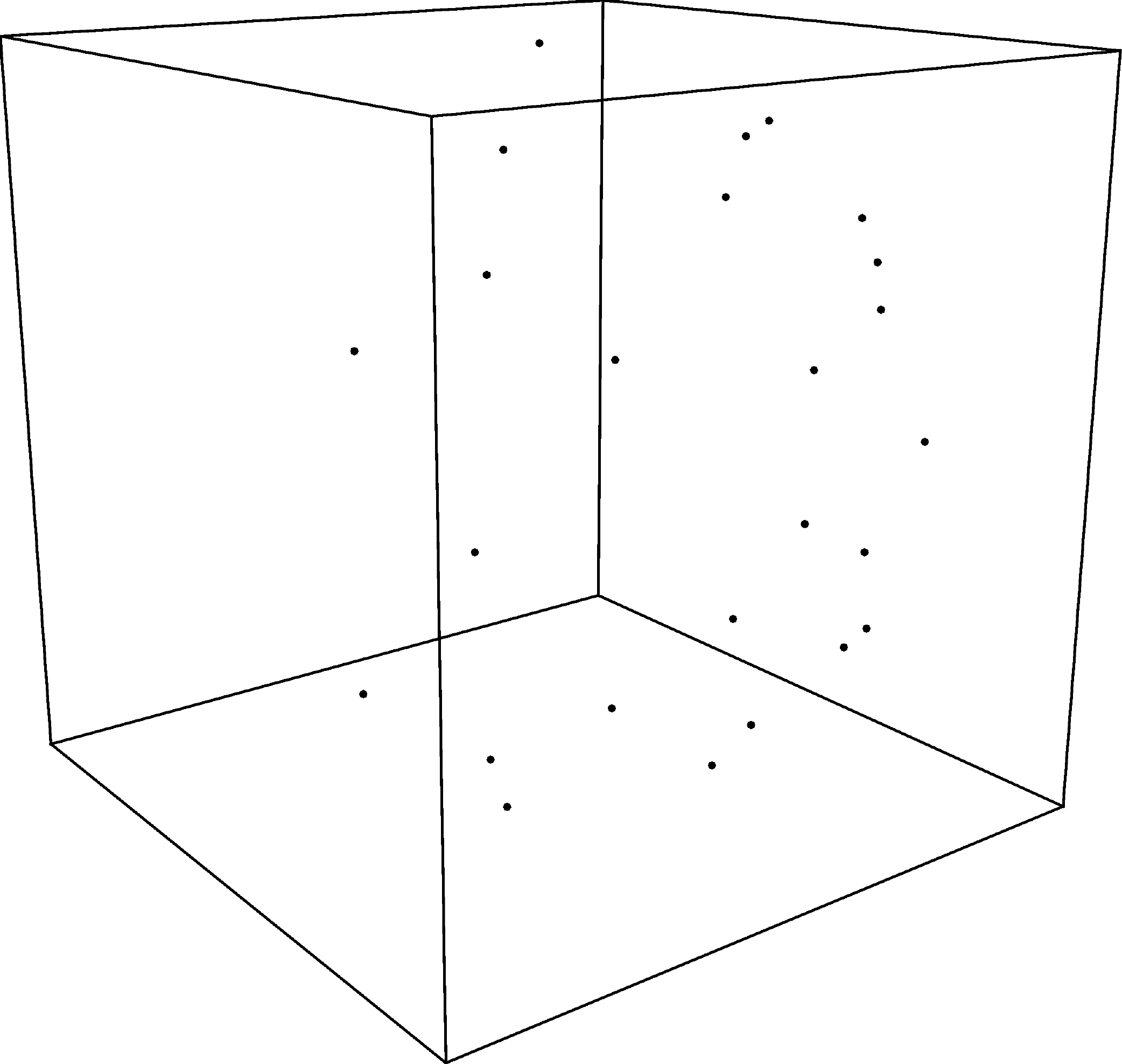}} \\
		\subfloat[]{\includegraphics[width=0.2\textwidth]{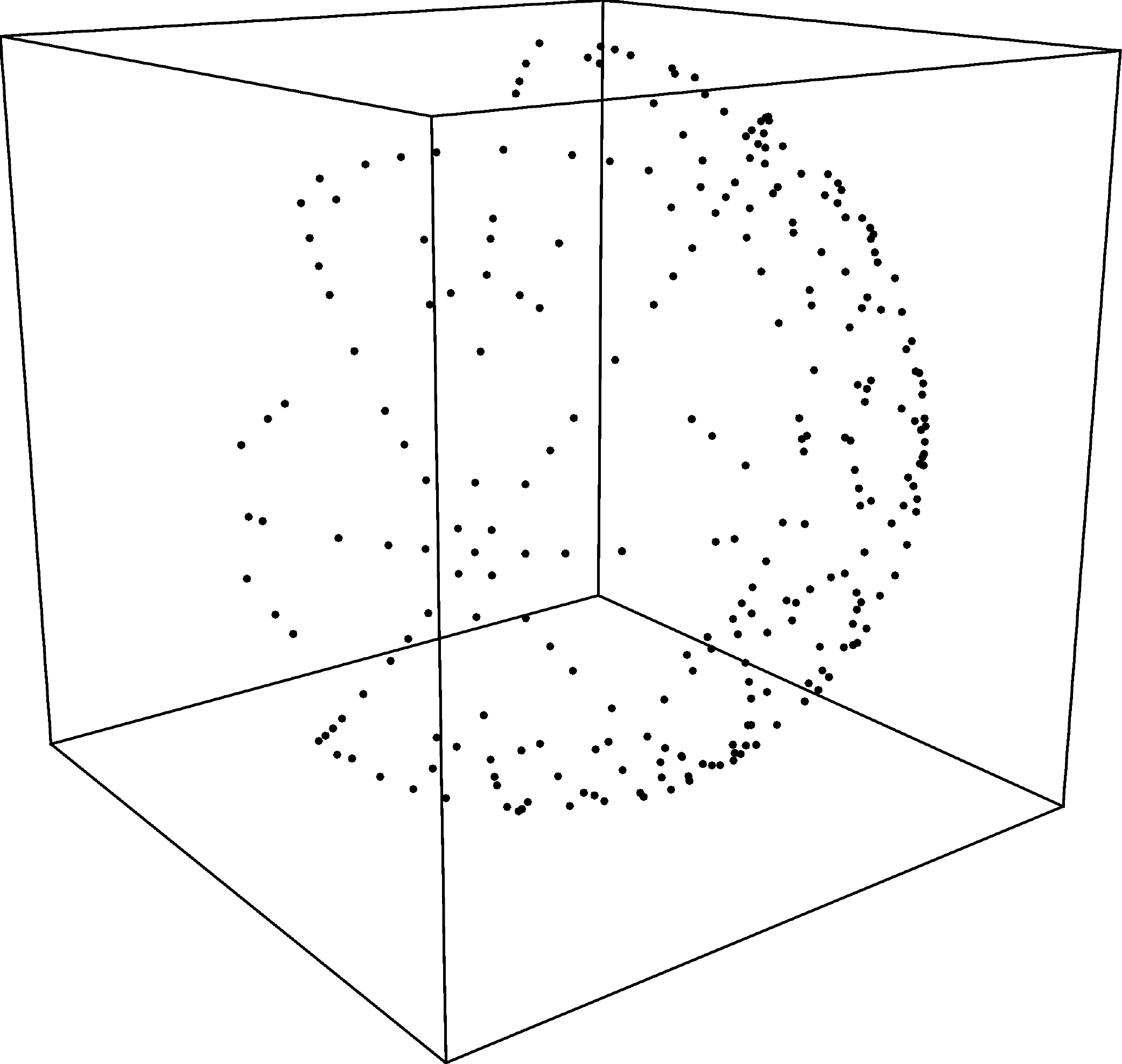}} & \subfloat[]{\includegraphics[width=0.2\textwidth]{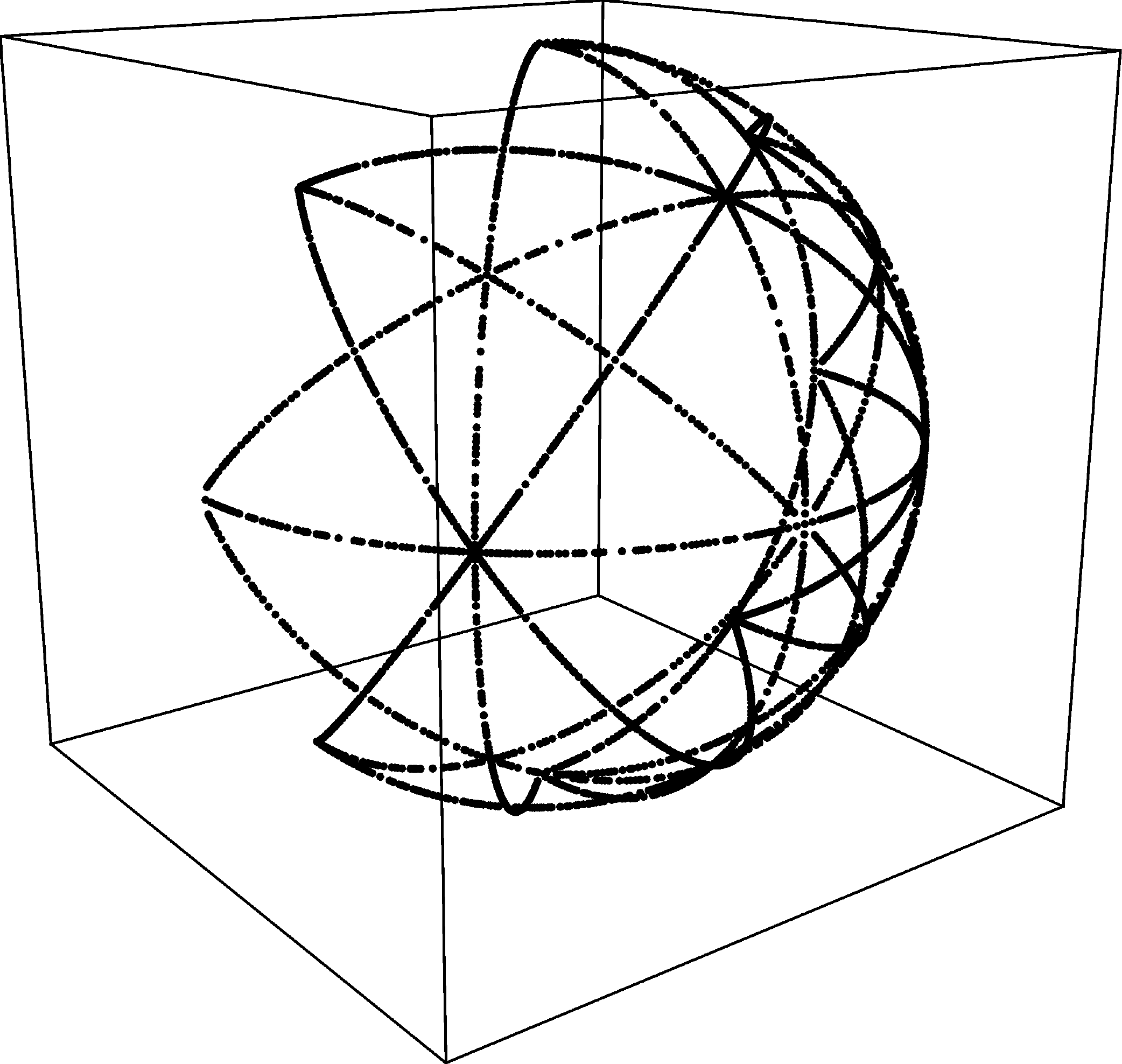}}\\
		\multicolumn{2}{c}{\subfloat[]{\includegraphics[width=0.46\textwidth]{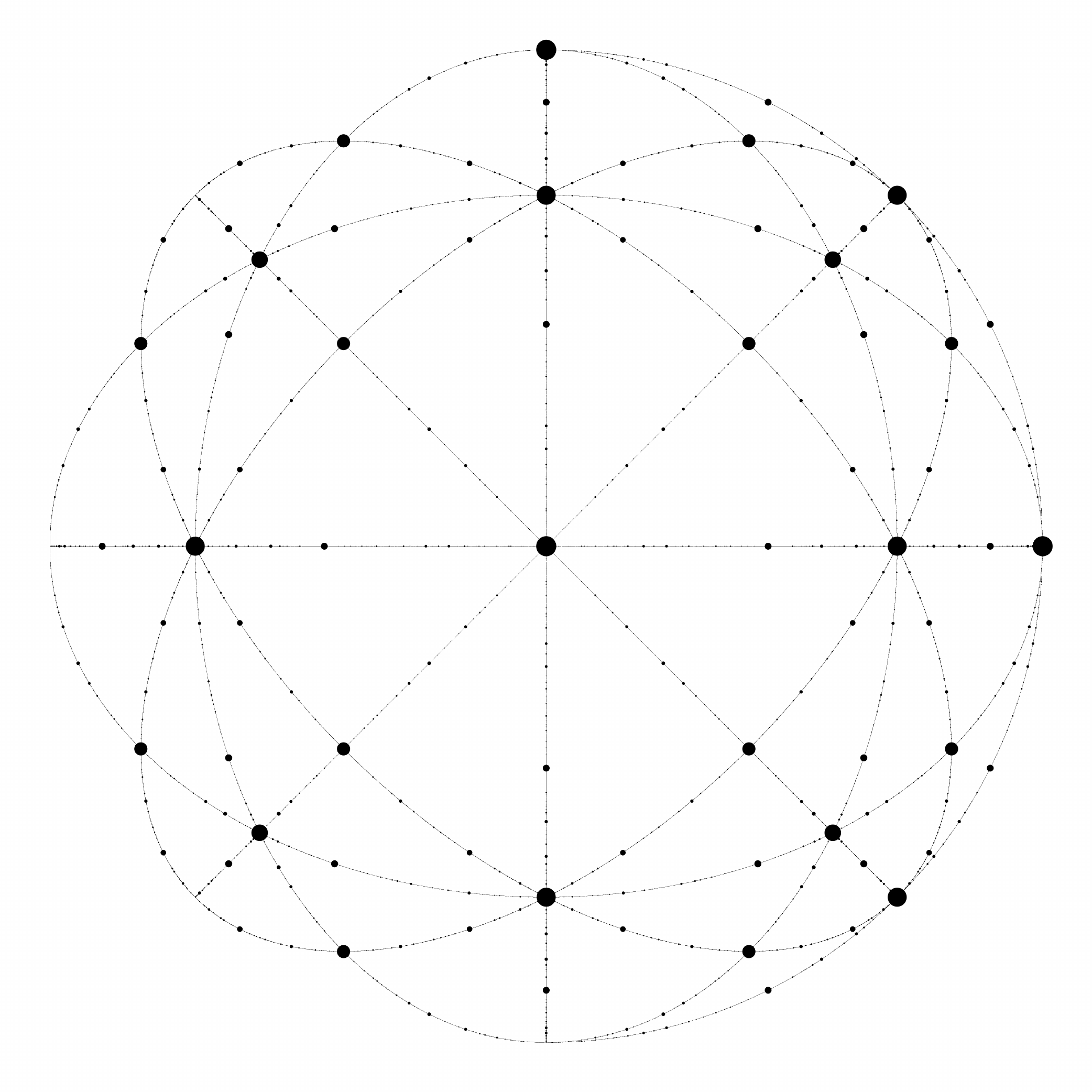}}}
	\end{tabular}
	\caption{Assuming that the experiment starts with a qutrit in one of the 13 quantum states of the Yu-Oh set, represented by the dots over a semisphere in (a), the successive figures show the possible postmeasurement quantum states after one (b), three (c), five (d), and seven measurements (e). The number of possible postmeasurement states is 25, 265, 3649, and 50293, respectively. All the states lie in one of the 13 semicircles corresponding to the states with real components orthogonal to the 13 states of the Yu-Oh set. However, not all of them occur with the same probability. To illustrate this, the volume of each point in (e) is proportional to the probability with which the corresponding state appears. Collectively, these figures exemplify the typical behavior of QSIC sets (e.g., \cite{CEG96,LBPC14}).}
	\label{fig:dis}
\end{figure}

When one uses the Yu-Oh set for unlimited 
sequential measurements on a single system, if at any point the system is in one of 
the 13 pure states of the Yu-Oh set and one measures one randomly chosen projector 
onto one of these 13 states, then the number of 
possible postmeasurement states does not remain constant but grows with the number 
of sequential measurements. In fact, some states are more probable than others 
(see Fig.~\ref{fig:dis}). This contrasts with the case of the Peres-Mermin set, 
where the number of possible postmeasurement states is constant and all states are equally probable.


{\em The notion of simulation and the relation to previous works.---}When talking about a classical simulation of a temporal process, it is important to specify what precisely shall be simulated and which conditions a simulation apparatus should meet. A general strategy for simulating temporal correlations is to use hidden Markov models
(HMMs) \cite{RJ86} or, when deterministic effects are considered, Mealy machines \cite{Mealy55}. There, the simulation apparatus is always in a definite internal state $k$, and for each internal state $k$, there is an output mechanism (e.g., a table $R_k$ containing all the results of the potential measurements) and an update mechanism (e.g., a table $U_k$ that describes the change of the internal state depending on the measurement). In such a model, however, it can easily happen that the simulation apparatus contains information about the future that cannot be derived from the past. By this we mean the following: consider two persons, where the first one only knows
the current internal state of the machine and the second one only knows the
past observation of measurements and results. Clearly, if the simulation
apparatus simulates all the correlations properly, the first person can
predict the future as well as the second person. For many processes, however,
it can happen that the first person can predict the future better, e.g., if
the given internal state $k$ predicts a deterministic outcome for the next
measurement, which cannot be deduced from the past. In this way, a simulation
apparatus can contain oracular information (i.e., information that cannot be obtained from the past) \cite{CEJM10}.

For our simulation, we restrict our attention to a simulation without oracular information. This leads to the notion of causal models and, more specifically, to $\varepsilon$ transducers, as explained below. These are also so-called unifilar processes, meaning that they are special HMMs, where the output derived from the internal state $k$ determines the update of the internal state. We note that with more general HMMs the memory required for the simulation can sometimes be reduced \cite{CEJM10, LA08} and that such models have been used to simulate the Peres-Mermin set \cite{K11, FK16}. Our restriction to causal models, however, is physically motivated by the demand that only the past observations should be used for simulating the future.


{\em Tools.---}To calculate the minimum memory that a classical system must have, a key observation is that our ideal experiments are examples of stochastic input-output processes that can be analyzed in information-theoretic terms. A stochastic process $\overleftrightarrow{{\cal Y}}$ is a one-dimensional chain $\ldots,Y_{-2},Y_{-1},Y_0,Y_1,Y_2,\ldots$ of discrete random variables $\{Y_t\}_{t \in \mathbb{Z}}$ that take values $\{y_t\}_{t \in \mathbb{Z}}$ over a finite or countably infinite alphabet ${\cal Y}$. An input-output process $\overleftrightarrow{Y}|\overleftrightarrow{X}$ with input alphabet ${\cal X}$ and output alphabet ${\cal Y}$ is a collection of stochastic processes $\overleftrightarrow{Y}|\overleftrightarrow{X} \equiv \{ \overleftrightarrow{Y} | \overleftrightarrow{x} \}_{\overleftrightarrow{x} \in \overleftrightarrow{{\cal X}}}$, where each such process $\overleftrightarrow{Y}|\overleftrightarrow{x}$ corresponds to all possible output sequences $\overleftrightarrow{Y}$ given a particular bi-infinite input sequence $\overleftrightarrow{x}$. It can be represented as a finite-state automaton or, equivalently, as a hidden Markov process. In our experiment, $x_t$ is the observable measured at time $t$ and $y_t$ is the 
corresponding outcome. By $\overleftarrow{X}$ we denote the chain of previous measurements, $\ldots,X_{t-2},X_{t-1}$, by $\overrightarrow{X}$ we denote $X_t,X_{t+1},\ldots$, and by $\overleftrightarrow{X}$ we denote the chain $\ldots,X_{t-1},X_{t},X_{t+1},\ldots$. Similarly, $\overleftarrow{Y}$, $\overrightarrow{Y}$, and $\overleftrightarrow{Y}$ denote the past, future, and all outcomes, respectively, while $\overleftarrow{Z}$, $\overrightarrow{Z}$, and $\overleftrightarrow{Z}$ denote the past, future, and all pairs of measurements and outcomes. For deriving physical consequences, we have to consider the minimal and optimal representation of this process.

As proven in Ref.~\cite{BC14}, the fact that each of our experiments is an input-output process implies that for each of them there exists a unique minimal and optimal predictor of the process, i.e., a unique finite-state machine with minimal entropy over the state probability distribution and maximal mutual information with the process's future output given the process's input-output past and the process's future input. This machine is called the process's $\varepsilon$ transducer \cite{BC14} and is the extension of the so-called $\varepsilon$ machines \cite{CY89, SC01}. An $\varepsilon$ transducer of an input-output process is a tuple $({\cal X}, {\cal Y}, {\cal S}, {\cal T})$ consisting of the process's input and output alphabets ${\cal X}$ and ${\cal Y}$, the set of causal states ${\cal S}$, and the set of corresponding conditional transition probabilities ${\cal T}$. The causal states $s_{t-1} \in {\cal S}$ are the equivalence classes in which the set of input-output pasts $\overleftarrow{{\cal Z}}$ can be partitioned in such a way that two input-output pasts $\overleftarrow{z}$ and $\overleftarrow{z}'$ are equivalent if and only if the probabilities $P(\overrightarrow{Y}|\overrightarrow{X},\overleftarrow{Z}=\overleftarrow{z})$ and $P(\overrightarrow{Y}|\overrightarrow{X},\overleftarrow{Z}=\overleftarrow{z}')$ are equal. The causal states are a so-called sufficient statistic of the process. They store all the information about the past needed to predict the output and as little as possible of the remaining information overhead contained in the past. The Shannon entropy over the stationary distribution of the causal states $H({\cal S})$ is the so-called statistical complexity and represents the minimum internal entropy needed to be stored to optimally compute future measurement outcomes (this quantity generally depends on how our measurements $\overleftrightarrow{X}$ are selected; here, we assume each $X_t$ is selected from a uniform probability distribution). The set of conditional transition probabilities ${\cal T} \equiv \{P(S_{t+1} = s_j,Y_t=y|S_t=s_i,X_t=x)\}$ governs the evolution.


\begin{figure}[tb]
	\centering
	\includegraphics[width=7.2cm]{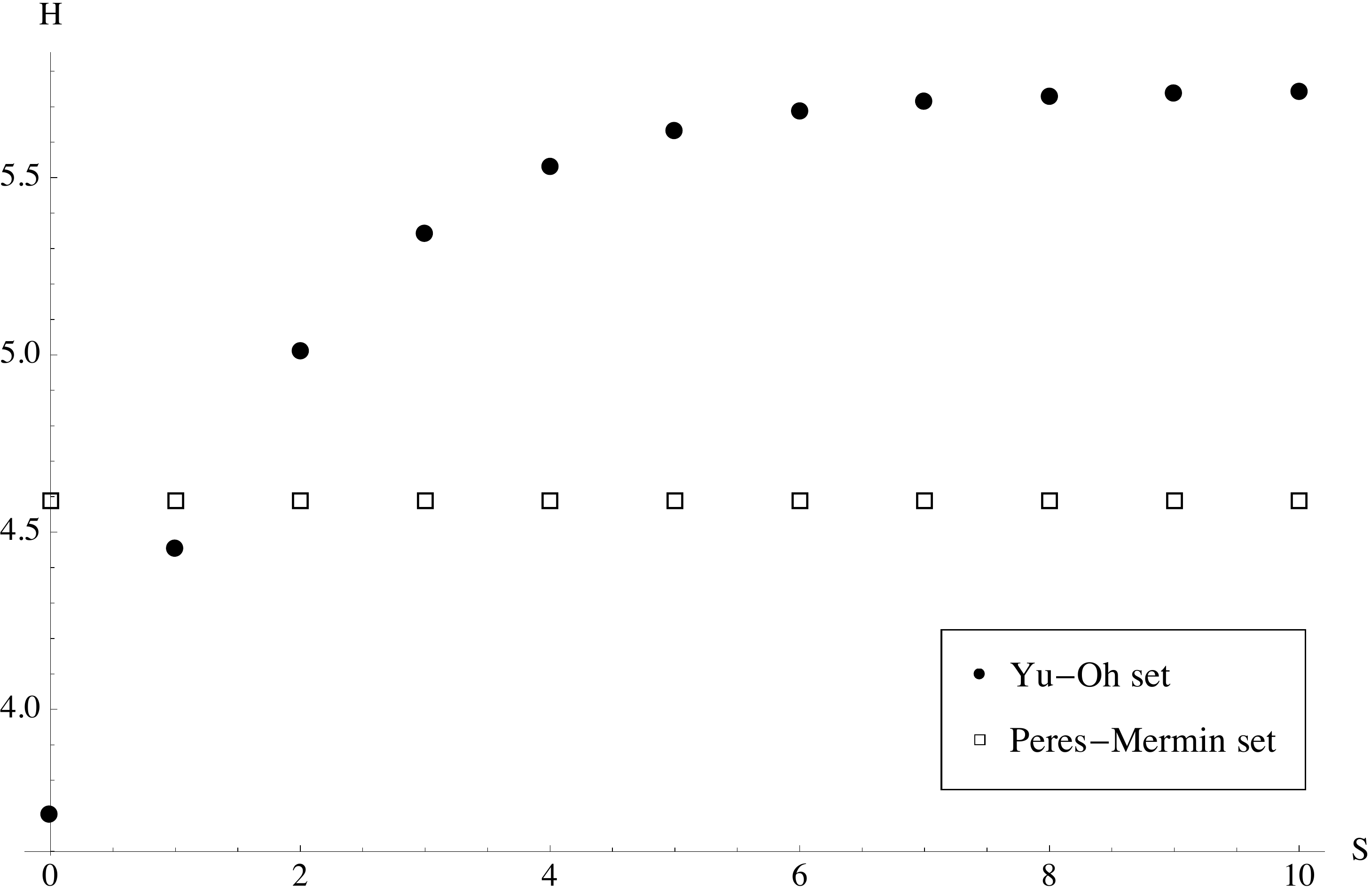}
	\caption{Classical memory in bits needed to simulate a sequential 
	Yu-Oh and Peres-Mermin experiments, as given by Eq.\ (\ref{main}), as a function of 
	the number of steps, as defined in the caption of Fig.\ \ref{fig:dis} for the case of Yu-Oh (and similarly for the case of Peres-Mermin). 
	Values are obtained from considering all possible measurement sequences 
	of a given number of steps and then analytically calculating the corresponding 
	results and postmeasurement states. \label{Fig4}}
	\label{fig:step10entropy}
\end{figure}


{\em Minimum memory needed to simulate QSIC.---}The $\varepsilon$ transducers of associated with the QSIC experiments 
have a particular property, namely, that there is a one-to-one correspondence between 
causal states $s_t$ and quantum states $|\psi_{t}\rangle \in \Phi$, where $\Phi$ is the
set of possible states occurring after a measurement (for completeness, a proof is presented in the Supplemental Material). Therefore, 
the minimum number of bits a finite-state classical machine must have to simulate the 
predictions of quantum theory for a QSIC experiment with unlimited sequential measurements 
chosen uniformly at random is given by the Shannon entropy
\begin{equation}
H=-\sum_i p_i \log_2 p_i.
\label{main}
\end{equation}
In \eqref{main}, $p_i$ is the probability of each quantum state achievable during the 
experiment's occurrence and, in general, depends on the 
distribution in which different measurements are chosen.

For the Peres-Mermin set, there are 24 causal states, each occurring with 
equal probability (see Fig.~\ref{Fig2}). Hence, a simulation with an 
$\varepsilon$ transducer requires $\log_2(24)=4.585$ bits to imitate a 
quantum system of 2 qubits. This classical memory is significantly 
higher than the classical information-carrying-capacity of the quantum 
system that produces these correlations.

For the Yu-Oh set, the calculations are more involved. The reason is that 
the longer the measurement sequence is, the more possible quantum states 
can occur as postmeasurement states. In addition, the quantum states do 
not occur with the same probability; see Fig.~\ref{fig:dis}. For small 
sequences up to length ten, however, all the states and probabilities 
can be analytically computed. The results imply that if only the last 
ten measurements and results are included, at least $5.740$ bits are
required for the simulation (see Fig.~\ref{Fig4}).

A proper comparison with the amount of memory required to simulate noncontextual sets is obtained by noticing that the memory required to reproduce the predictions of quantum mechanics when we restrict the measurements 
to subsets (of the QSIC sets) that cannot produce contextuality is $2$ bits for the Peres-Mermin set and $\log_2 3 \approx 1.585$ bits for the Yu-Oh set. These values are obtained as follows. Contextuality is an impossibility of a joint probability distribution over a single probability space. For sequences of projective measurements, incompatibility implies the nonexistence of a joint probability distribution. Therefore, the memory needed to simulate noncontextual sets is the one required to reproduce the predictions of quantum mechanics for subsets of mutually compatible measurements of the QSIC set, which is $\log_2 d$ bits for any QSIC set of dimension $d$. Notice that contextuality requires incompatibility, but also that measurements can be grouped into mutually compatible subsets so that each measurement belongs to at least two of them. Therefore, simulating a set of incompatible measurements not restricted by these rules may require more memory.

One might conjecture that the minimal memory necessary to classically simulate QSIC must be related to the degree of contextuality. However, the relation is difficult to trace. For example, while the minimal memory necessary to simulate classically the Peres-Mermin and Yu-Oh sets is larger for Yu-Oh, the degree of contextuality that can be measured by, e.g., the ratio between the violation and the noncontextual bound for the optimal noncontextuality inequalities \cite{KBLGC12} is $1.5$ for Peres-Mermin and $1.107$ for Yu-Oh, showing that contextuality is higher for Peres-Mermin. The same conclusion can be reached by adopting other measures of contextuality \cite{AB11,GHHHJKW14}. Therefore, understanding the connection between memory and the degree of contextuality is an interesting open problem that should be addressed in the future. Here, also the effects of noise and
imperfections should be considered.


{\em Conclusions.---}The question of which classical resources are needed
for simulating quantum effects is central for the connection of the foundations 
of quantum theory with quantum information. By applying the tools of complexity science, we have shown how to calculate the amount of
memory a classical system would need to simulate quantum state-independent 
contextuality in the case of a single quantum system submitted to an infinite 
sequence of measurements randomly chosen from any finite set. 
Our result precisely quantifies the quantum vs classical advantage of a phenomenon, quantum state-independent contextuality, discovered 50 years ago and shows how profitable may be combining previously unrelated disciplines, such as complexity and quantum information.

Our result opens a way to test systems for their quantumness.
Suppose we have a system whose internal functioning is unknown
and that is submitted to sequential measurements for which a
classical simulation requires more memory than the one allowed
by the Bekenstein bound. Here, the Bekenstein bound refers to
the limit on the entropy that can be contained in a physical
system with given size and energy \cite{Bekenstein72}. We may assume that no
system can store and process information beyond the Bekenstein
bound and can test whether the system is not emitting heat
due to Landauer's principle (which states that the erasure of
classical information implies some heat emission \cite{Landauer61}). If
this heat is not found, then our result allows us to certify
that the system is in fact quantum and not a classical
simulation. Therefore, we can use its quantum features for
information processing.

On the other hand, our result could also inspire new techniques in complexity science, where there is a growing interest in the value of quantum theory for simulating otherwise difficult to simulate classical processes (e.g., \cite{TGVG17,PGHWP17}). In this respect, our result could pinpoint the properties of classical processes that make them particularly amenable to improved modeling using quantum systems and thus also further catalyze the use of quantum methods in complexity science.

\medskip


\begin{acknowledgments} 
	We thank Matthias Kleinmann, Jan-{\AA}ke Larsson, and Karoline Wiesner for discussions, and Jayne Thompson for help in the Supplemental Material. This work was supported by Project No.\ FIS2014-60843-P, ``Advanced Quantum Information'' (MINECO, Spain) with FEDER funds, the project ``Photonic Quantum Information'' (Knut and Alice Wallenberg Foundation, Sweden), and the Singapore National Research Foundation Fellowship No.\ NRF-NRFF2016-02. A.C.\ is also supported by the FQXi Large Grant ``The Observer Observed: A Bayesian Route to the Reconstruction of Quantum Theory.'' M.G.\ is also supported by the John Templeton Foundation Grant No.\ 53914 ``Occam's Quantum Mechanical Razor: Can Quantum Theory Admit the Simplest Understanding of Reality?'' and the FQXi Large Grant ``Observer-Dependent Complexity: The Quantum-Classical Divergence over `What is Complex?'\,'' O.G.\ is also supported by the DFG and the ERC (Consolidator Grant No.\ 683107/TempoQ). Z.-P.X.\ is also supported by the Natural Science Foundation of China (Grant No.\ 11475089) and the China Scholarship Council.
\end{acknowledgments}



\section*{Supplemental Material}


The $\varepsilon$-transducers associated to quantum state-independent contextuality (QSIC) experiments have a particular property, namely, that there is a one-to-one correspondence between causal states $s_t$ and quantum states $|\psi_{t}\rangle \in \Phi$, where $\Phi$ is the
set of possible states occurring after a measurement. This can be seen as follows.

Consider a quantum system measured at discrete time steps $t$ by some chosen rank-$1$ projector $M_t \in \mathcal{X}$ with measurement outcome  $y_t \in \{0,1\}$. Let the input-output process $\overleftrightarrow{Y}|\overleftrightarrow{X}$, with input alphabet  $\mathcal{X}$ and output alphabet $\mathcal{Y} = \{0,1\}$, be the  input-output process associated to the experiment. We define the experiment to be causally complete if it satisfies that: (a) Each measurement choice $M_t$ is made independently, such that $P(M_t = x)$ is finite for each $x \in \mathcal{X}$. (b) The input-output past of the system is sufficient to ascertain exactly the state the system is in. (c) Any two different quantum states of the system attained during the experiment can be distinguished statistically by measurements in $\mathcal{X}$.

Note that (a) is always true when measurements are chosen at random. For finite QSIC sets (b) is generically true, i.e., the probability for a past that does not determine the actual state decreases exponentially with the number of steps. (c) holds for the Yu-Oh and Peres-Mermin sets and also for any QSIC set made of rank-$1$ projectors, and any QSIC set can always be implemented using a QSIC set with only rank-$1$ projectors. The fact that (c) holds for any QSIC set made of rank-$1$ projectors follows from the fact that any state of a QSIC set belongs to, al least, two different basis, and from the fact  the attained states  are orthogonal to some state of the QSIC set (see, e.g., Fig.\ 3 in the main text). Then, to distinguish between two attained states $|\psi_{a}\rangle$ and $|\psi_{b}\rangle$ that are not orthogonal to the same state of the QSIC set, one can measure the projection on one state of the QSIC set orthogonal to $|\psi_{a}\rangle$ but not to $|\psi_{b}\rangle$. To distinguish between two attained states that are both orthogonal to the same subspace span by states of the QSIC set, one can measure the projection on one state of the QSIC set orthogonal to that subspace. We can now establish the following:


\begin{theorem} 
Let the input-output process $\overleftrightarrow{Y}|\overleftrightarrow{X}$ 
with input alphabet ${\cal X}$ and output alphabet ${\cal Y}$ be one that 
describes the input-output behavior of a causally complete measurement 
procedure. Let $\Phi = \{\ket{\phi_i}\}$ be the set of possible states that the quantum
system can take after the measurements. Then, $\Phi$ is in one-to-one 
correspondence with the causal states of 
$\overleftrightarrow{Y}|\overleftrightarrow{X}$.
	
	
\begin{proof} 
Observe that conditions (a) and (b) guarantee that observation 
of each possible past $\overleftarrow{z} = (\overleftarrow{x}, \overleftarrow{y})$ 
implies that the actual state can be deduced from the past. Thus, the state of 
the quantum system in the present does not contain any oracular information -- 
it is possible to determine the state at $t = 0$ entirely by looking at past 
input-output behavior. Thus the encoding function 
$s(\overleftarrow{z}) = \phi_{\overleftarrow{z}} \in \Phi$ that maps each possible 
past to the resulting state of the system in the present exists. Therefore, to 
establish $\Phi$ is in one-to-one correspondence with the causal states, we 
need only to show that $s(\overleftarrow{z}) = s(\overleftarrow{z}')$ if and 
only if they are of coinciding future input-output behavior, i.e., 
$P(\overrightarrow{Y} | \overrightarrow{X}, \overleftarrow{Z} = \overleftarrow{z}) 
= P(\overrightarrow{Y} | \overrightarrow{X}, \overleftarrow{Z} = \overleftarrow{z}')$. 
In both directions we will use proof by contrapositive.
		
To prove the forward direction, note that the state of the quantum system uniquely determines the distribution over future outcomes $\overrightarrow{Y}$, for any potential measurement sequence $\overrightarrow{x}$. Thus whenever $P(\overrightarrow{Y} | \overrightarrow{x}, \overleftarrow{z}) \neq P(\overrightarrow{Y} | \overrightarrow{x}, \overleftarrow{z}')$ for some $\overrightarrow{x}$, we must have $\phi_{\overleftarrow{z}} \neq \phi_{\overleftarrow{z}'}$.
For the reverse direction note that by condition (c), the set of measurements $M_x \in \mathcal{X}$ allows us to distinguish statistically any state. Thus, if $\phi_{\overleftarrow{z}} \neq \phi_{\overleftarrow{z}'}$, then there exists $M \in \mathcal{M}$ such that $\rm{tr} (\rho_{\overleftarrow{z}} M ) \neq \rm{tr} (\phi_{\overleftarrow{z}'}M)$. Thus $\rho_{\overleftarrow{z}} \neq \rho_{\overleftarrow{z}'}$ implies $P(Y_0 | X_0 = M, \overleftarrow{z}) \neq P(Y_0 | X_0 = M, \overleftarrow{z}')$.
\end{proof}
\end{theorem}



\begin{thebibliography}{99}


\bibitem{Bell64}
J. S. Bell,
On the Einstein Podolsky Rosen paradox,
Physics \textbf{1}, 195 (1964).


\bibitem{BCT99}
G. Brassard, R. Cleve, and A. Tapp,
Cost of Exactly Simulating Quantum Entanglement with Classical Communication,
\href{http://dx.doi.org/10.1103/PhysRevLett.83.1874}{Phys. Rev. Lett. \textbf{83}, 1874 (1999).}

\bibitem{Steiner00}
M. Steiner,
Towards quantifying non-local information transfer: finite-bit non-locality,
\href{http://dx.doi.org/10.1016/S0375-9601(00)00315-7}{Phys. Lett. A \textbf{270}, 239 (2000).}

\bibitem{CGM00}
N. J. Cerf, N. Gisin, and S. Massar,
Classical Teleportation of a Quantum Bit,
\href{https://doi.org/10.1103/PhysRevLett.84.2521}{Phys. Rev. Lett. \textbf{84}, 2521 (2000).}

\bibitem{BT03}
D. Bacon and B. F. Toner,
Bell Inequalities with Auxiliary Communication,
\href{https://doi.org/10.1103/PhysRevLett.90.157904}{Phys. Rev. Lett. \textbf{90}, 157904 (2003).}

\bibitem{TB03}
B. F. Toner and D. Bacon,
Communication Cost of Simulating Bell Correlations,
\href{https://doi.org/10.1103/PhysRevLett.91.187904}{Phys. Rev. Lett. \textbf{91}, 187904 (2003).}

\bibitem{Pironio03}
S. Pironio,
Violations of Bell inequalities as lower bounds on the communication cost of nonlocal correlations,
\href{https://doi.org/10.1103/PhysRevA.68.062102}{Phys. Rev. A \textbf{68}, 062102 (2003).}

\bibitem{CGMP05}
N. J. Cerf, N. Gisin, S. Massar, and S. Popescu,
Simulating Maximal Quantum Entanglement without Communication,
\href{http://dx.doi.org/10.1103/PhysRevLett.94.220403}{Phys. Rev. Lett. \textbf{94}, 220403 (2005).}


\bibitem{Bell66}
J. B. Bell,
On the problem of hidden Variables in quantum mechanics,
\href{http://dx.doi.org/10.1103/RevModPhys.38.447}{Rev. Mod. Phys. \textbf{38}, 447 (1966).}

\bibitem{KS67}
S. Kochen and E. P. Specker,
The problem of hidden variables in quantum mechanics,
\href{http://www.jstor.org/stable/24902153}{J. Math. Mech. \textbf{17}, 59 (1967).}

\bibitem{KCBS08}
A. A. Klyachko, M. A. Can, S. Binicio\u{g}lu, and A. S. Shumovsky,
Simple Test for Hidden Variables in Spin-1 Systems,
\href{http://dx.doi.org/10.1103/PhysRevLett.101.020403}{Phys. Rev. Lett. \textbf{101}, 020403 (2008).}

\bibitem{Cabello08}
A. Cabello,
Experimentally Testable State-Independent Quantum Contextuality,
\href{http://dx.doi.org/10.1103/PhysRevLett.101.210401}{Phys. Rev. Lett. \textbf{101}, 210401 (2008).}

\bibitem{BBCP09}
P. Badzi\c{a}g, I. Bengtsson, A. Cabello, and I. Pitowsky,
Universality of State-Independent Violation of Correlation Inequalities for Noncontextual Theories,
\href{http://dx.doi.org/10.1103/PhysRevLett.103.050401}{Phys. Rev. Lett. \textbf{103}, 050401 (2009).}

\bibitem{YO12}
S. Yu and C. H. Oh,
State-Independent Proof of Kochen-Specker Theorem with 13 Rays,
\href{http://dx.doi.org/10.1103/PhysRevLett.108.030402}{Phys. Rev. Lett. \textbf{108}, 030402 (2012).}

\bibitem{KBLGC12}
M. Kleinmann, C. Budroni, J.-\AA. Larsson, O. G{\"u}hne, and A. Cabello,
Optimal Inequalities for State-Independent Contextuality,
\href{http://dx.doi.org/10.1103/PhysRevLett.109.250402}{Phys. Rev. Lett. \textbf{109}, 250402 (2012).}


\bibitem{K11}
 M. Kleinmann, O. G{\"u}hne, J. R Portillo, J.-\AA. Larsson, and A. Cabello,
 Memory cost of quantum contextuality,
 \href{http://dx.doi.org/10.1088/1367-2630/13/11/113011}{New J. Phys. \textbf{13}, 113011 (2011).}

\bibitem{Cabello12}
A. Cabello,
The role of bounded memory in the foundations of quantum mechanics,
\href{http://dx.doi.org/10.1007/s10701-010-9507-2}{Found. Phys. \textbf{42}, 68 (2012).}

\bibitem{FK16}
 G. Fagundes and M. Kleinmann,
 Memory cost for simulating all quantum correlations of the Peres-Mermin scenario,
 \href{https://doi.org/10.1088/1751-8121/aa7ab3}{J. Phys. A {\bf 50}, 325302 (2017).}


\bibitem{WLK16}
 M. Wajs, S.-Y. Lee, P. Kurzy\'{n}ski, and D. Kaszlikowski,
 State-recycling method for testing quantum contextuality,
 \href{http://dx.doi.org/10.1103/PhysRevA.93.052104}{Phys. Rev. A \textbf{93}, 052104 (2016).}

\bibitem{LMZNACH17}
F. M. Leupold, M. Malinowski, C. Zhang, V. Negnevitsky, J. Alonso, A. Cabello, and J. P. Home,
Sustained state-independent quantum contextual correlations from a single ion,
\href{https://arxiv.org/abs/1706.07370}{\eprint{arXiv:1706.07370}.}

\bibitem{CKB15}
A. Cabello, M. Kleinmann, and C. Budroni,
Necessary and Sufficient Condition for Quantum State-Independent Contextuality,
\href{http://dx.doi.org/10.1103/PhysRevLett.114.250402}{Phys. Rev. Lett. \textbf{114}, 250402 (2015).}

\bibitem{CKP16}
A. Cabello, M. Kleinmann, and J. R. Portillo,
Quantum state-independent contextuality requires 13 rays,
\href{http://dx.doi.org/10.1088/1751-8113/49/38/38LT01}{J. Phys. A \textbf{49}, 38LT01 (2016).}


\bibitem{Peres90}
A. Peres,
Incompatible results of quantum measurements,
\href{http://dx.doi.org/10.1016/0375-9601(90)90172-K}{Phys. Lett. A \textbf{151}, 107 (1990).}

\bibitem{Mermin90b}
N. D. Mermin,
Simple Unified Form for the Major No-Hidden-Variables Theorems,
\href{http://dx.doi.org/10.1103/PhysRevLett.65.3373}{Phys. Rev. Lett. \textbf{65}, 3373 (1990).}

\bibitem{KZG09}
G. Kirchmair, F. Z\"ahringer, R. Gerritsma, M. Kleinmann, O. G{\"u}hne, A. Cabello, R. Blatt, and C. F. Roos,
State-independent experimental test of quantum contextuality,
\href{http://doi:10.1038/nature08172}{Nature (London) \textbf{460}, 494 (2009).}

\bibitem{ARBC09}
E. Amselem, M. R{\aa }dmark, M. Bourennane, and A. Cabello,
State-Independent Quantum Contextuality with Single Photons,
\href{http://dx.doi.org/10.1103/PhysRevLett.103.160405}{Phys. Rev. Lett. \textbf{103}, 160405 (2009).}

\bibitem{LHC16}
B.-H. Liu, X.-M. Hu, J.-S. Chen, Y.-F. Huang, Y.-J. Han, C.-F. Li, G.-C. Guo, and A. Cabello,
Nonlocality from Local Contextuality,
\href{https://doi.org/10.1103/PhysRevLett.117.220402}{Phys. Rev. Lett. \textbf{117}, 220402 (2016).}


\bibitem{Peres91}
A. Peres,
Two simple proofs of the Kochen-Specker theorem,
\href{https://doi.org/10.1088/0305-4470/24/4/003}{J. Phys. A \textbf{24}, L175 (1991).}

\bibitem{ZWDCLHYD12}
C. Zu, Y.-X. Wang, D.-L. Deng, X.-Y. Chang, K. Liu, P.-Y. Hou, H.-X. Yang, and L.-M. Duan,
State-Independent Experimental Test of Quantum Contextuality in an Indivisible System,
\href{http://dx.doi.org/10.1103/PhysRevLett.109.150401}{Phys. Rev. Lett. \textbf{109}, 150401 (2012).}

\bibitem{ABB13}
E. Amselem, M. Bourennane, C. Budroni, A. Cabello, O. G\"uhne, M. Kleinmann, J.-\AA. Larsson, and M. Wie\'{s}niak,
Comment on ``State-Independent Experimental Test of Quantum Contextuality in an Indivisible System'',
\href{https://doi.org/10.1103/PhysRevLett.110.078901}{Phys. Rev. Lett. \textbf{110}, 078901 (2013).}

\bibitem{ZUZ13}
X. Zhang, M. Um, J. Zhang, S. An, Y. Wang, D.-L. Deng, C. Shen, L.-M. Duan, and K. Kim,
State-Independent Experimental Test of Quantum Contextuality with a Single Trapped Ion,
\href{http://dx.doi.org/10.1103/PhysRevLett.110.070401}{Phys. Rev. Lett. \textbf{110}, 070401 (2013).}


\bibitem{CEG96}
A. Cabello, J. M. Estebaranz, and G. Garc\'{\i}a-Alcaine,
Bell-Kochen-Specker theorem: A proof with 18 vectors.
\href{http://dx.doi.org/10.1016/0375-9601(96)00134-X}{Phys. Lett. A \textbf{212}, 183 (1996).}

\bibitem{LBPC14}
P. Lison\v{e}k, P. Badzi\c{a}g, J. R. Portillo, and A. Cabello,
Kochen-Specker set with seven contexts,
\href{http://dx.doi.org/10.1103/PhysRevA.79.012102}{Phys. Rev. A \textbf{89}, 042101 (2014).}


\bibitem{RJ86}
L. R. Rabiner and B. H. Juang,
An introduction to hidden Markov models,
\href{http://doi.org/10.1109/MASSP.1986.1165342}{IEEE ASSP Mag. {\bf 3}, 4 (1986).}

\bibitem{Mealy55}
G. H. Mealy,
A method for synthesizing sequential circuits,
\href{http://doi.org/10.1002/j.1538-7305.1955.tb03788.x}{Bell Syst. Tech. J. \textbf{34}, 1045 (1955).}

\bibitem{CEJM10}
J. P. Crutchfield, C. J. Ellison, R. G. James, and J. R. Mahoney,
Synchronization and control in intrinsic and designed computation: An information-theoretic analysis of competing models of stochastic computation,
\href{http://dx.doi.org/10.1063/1.3489888}{Chaos \textbf{20}, 037105 (2010).}	

\bibitem{LA08}
W. L\"ohr and N. Ay,
in
{\em Complex Sciences. Complex 2009, Part I}, edited by J. Zhou,
Lecture Notes of the Institute for Computer Sciences, Social Informatics and Telecommunications Engineering Vol.\ 4 (Springer, Berlin, 2009), p.\ 265.


\bibitem{BC14}
N. Barnett and J. P. Crutchfield,
Computational mechanics of input-output processes: Structured transformations and the $\varepsilon$-transducer,
\href{http://dx.doi.org/10.1007%2Fs10955-015-1327-5}{J. Stat. Phys. \textbf{161}, 404 (2015).}
	
\bibitem{CY89}
J. P. Crutchfield and K. Young,
Inferring Statistical Complexity,
\href{http://dx.doi.org/10.1103/PhysRevLett.63.105}{Phys. Rev. Lett. \textbf{63}, 105 (1989).}
	
\bibitem{SC01}
C. R. Shalizi and J. P. Crutchfield,
Computational mechanics: Pattern and prediction, structure and simplicity,
\href{http://dx.doi.org/10.1023/A:1010388907793}{J. Stat. Phys. \textbf{104}, 817 (2001).}



\bibitem{AB11}
S. Abramsky and A. Brandenburger,
The sheaf-theoretic structure of non-locality and contextuality,
\href{https://doi.org/10.1088/1367-2630/13/11/113036}{New J. Phys. \textbf{13}, 113036 (2011).}

\bibitem{GHHHJKW14}
A. Grudka, K. Horodecki, M. Horodecki, P. Horodecki, R. Horodecki, P. Joshi, W. K{\l}obus, and A. W\'ojcik,
Quantifying Contextuality,
\href{https://doi.org/10.1103/PhysRevLett.112.120401}{Phys. Rev. Lett. \textbf{112}, 120401 (2014).}


\bibitem{Bekenstein72}
J. D. Bekenstein,
Black holes and the second law,
\href{https://doi.org/10.1007/BF02757029}{Lett. Nuovo Cimento \textbf{4}, 737 (1972).}

\bibitem{Landauer61}
R. Landauer,
Irreversibility and heat generation in the computing process,
\href{http://dx.doi.org/10.1147/rd.53.0183}{IBM J. Res. Dev. \textbf{5}, 183 (1961).}

\bibitem{TGVG17}
J. Thompson, A. J. P. Garner, V. Vedral, and M. Gu,
Using quantum theory to simplify input–output processes,
\href{https://doi.org/10.1038/s41534-016-0001-3}{npj Quantum Inf. \textbf{3}, 6 (2017).}

\bibitem{PGHWP17}
M. S. Palsson, M. Gu, J. Ho, H. M. Wiseman, and G. J. Pryde,
Experimentally modeling stochastic processes with less memory by the use of a quantum processor,
\href{https://doi.org/10.1126/sciadv.1601302}{Sci. Adv. \textbf{3}, e1601302 (2017).}


\end{thebibliography}
\end{document}